# Experimental realization of a broadband illusion optics device

Chao Li, Xiankun Meng, Xiao Liu, Fang Li, Guangyou Fang

*Key Laboratory of Microwave and Electromagnetic Radiation, Institute of Electronics, Chinese Academy of Sciences (IECAS), 100190 Beijing, China*

Huanyang Chen

*School of Physical Science and Technology, Soochow University, Suzhou, Jiangsu 215006, China*

C. T. Chan

*Department of Physics and William Mong Institute of Nano Science and Technology, Hong Kong University of Science and Technology, Clear Water Bay, Hong Kong, China*

**Abstract:**
We experimentally demonstrate the first metamaterial "illusion optics" device - an "invisible gateway" by using a transmission-line medium. The device contains an open channel that can block electromagnetic waves at a particular frequency range. We also demonstrate that such a device can work in a broad frequency range.

Transformation optics [1, 2] has paved the way for the rational design of conceptual devices which have unprecedented control over the propagation of electromagnetic waves [3, 4, 5, 6, 7, 8, 9, 10, 11, 12, 13, 14]. Recently, the combination of the complementary media concept [15] with the transformation optics technique [16] has motivated a series of illusion optics devices [17, 18, 19, 20], which can change the electromagnetic scattering cross section of an object to create optical illusions. However, such effects are limited to theoretical analysis and numerical simulations. Using a transmission-line medium, we realized the first experimental demonstration of an illusion optics device, an "invisible gateway" [21], which is an open channel that appears to be blocked for waves of a selected range of frequencies. The gateway is a two dimensional device for proof-of-principle purpose. We also measured the performance of the device and demonstrated its broadband functionality.

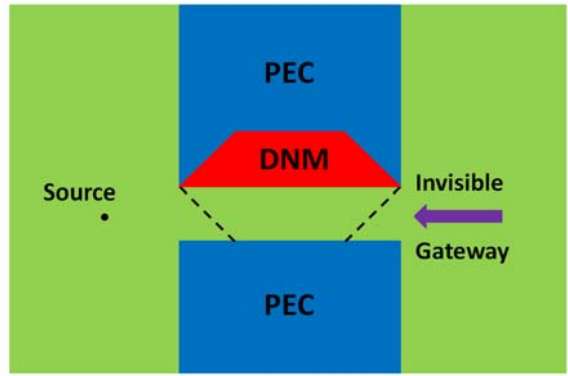

Figure 1 | Schematic picture of an electromagnetic invisible gateway. The air, PEC wall and DNM are denoted by green, blue and red colors respectively.

To start with, let us consider a configuration that is illustrated schematically in Figure 1. A perfect electric conductor (PEC, marked by blue color) wall partitions space (air, $\varepsilon=\mu=1$, denoted by green color) into two domains. A channel is then opened in the PEC wall with a double negative material (DNM, $\varepsilon'=\mu'=-1$) filled inside the trapezoidal region (marked by red color) and an air channel. From the viewpoint of transformation optics, the DNM will project the PEC boundary adjacent to it into another optically-equivalent PEC boundary (marked by the dashed line) so that the wall with an air channel will look like a continuous PEC wall for the observers outside at the designed frequency. In other words, the trapezoidal DNM slab can be viewed as a "complementary medium" which optically cancels positive index ($\varepsilon=\mu=1$) of the same thickness so that the open air channel is blocked at the frequency at which the DNM has $\varepsilon=\mu=-1$. Such an air channel was also called an "invisible gateway", meaning that an observer cannot "see" the open channel, but would rather see some reflection leading to the illusion that the channel is blocked. For the lack of a better term to describe the system, we will use "invisible gateway" hereafter. Electromagnetic (EM) wave incident from one side of the wall cannot reach to another side while other entities are allowed to pass through the open channel. Broadband functionality was also found in such a device in a previous publication. However, it is challenging to implement such a device electromagnetically due to the difficulty in realizing the DNM even for microwave frequencies. It should be noted that, such an invisible gateway can be applied to both two dimensional (2D) and three dimensional (3D) geometries. In this paper, we will only focus on the 2D case.

From the EM theory and the transmission-line (TL) theory, materials with specific permittivity and permeability tensors can be mimicked by periodic TL networks to give similar propagation behaviors. The TL versions of conventional double negative materials (DPMs) and artificial double negative materials (DNMs) have been investigated several years ago [22] and demonstrated experimentally to have the focusing effect that overcomes the classical diffraction limit [23].

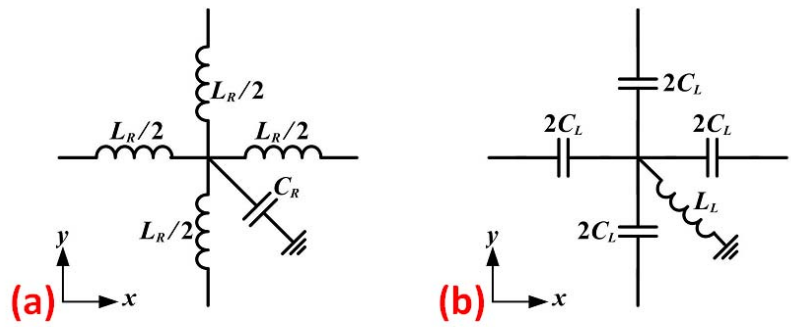

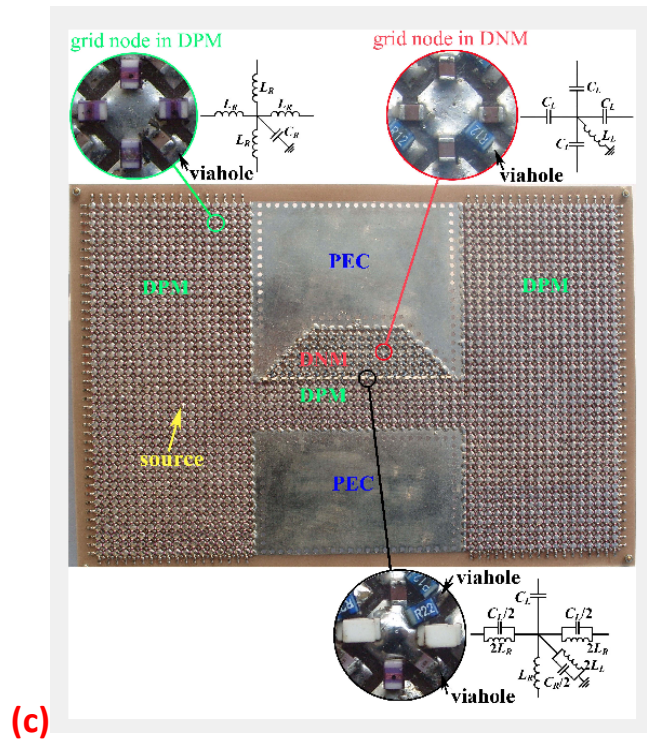

Figure 2 | Unit cells of periodic L-C structures to mimic the (a) DPMs and (b) DNMs, (c) the real experimental device.

Here, we choose a special kind of TL network - the periodic inductor-capacitor (L-C) network to mimic the configuration depicted in Figure 1. The unit cell to mimic the air region is shown in Figure 2a, where the series inductor and shunt capacitors can act as an isotropic medium with positive equal permittivity and permeability. The DNM is mimicked by the dual L-C configuration as shown in Figure 2b, where the position of L and C are interchanged, acting as a material with simultaneously negative values of permittivity and permeability.

There is a one-to-one mapping between the L-C network equations and the polarized transverse electric (or magnetic) Maxwell's equations with voltages V and currents I mapped to the field quantities E and H. In the long-wavelength limit, the dimension of the unit cell $\Delta$ is much smaller than the wavelength, therefore the relationship between the circuit parameters (the capacitance and inductance in the network) and the materials parameters (the effective permittivity and permeability) can be derived as

$$\varepsilon_R = \frac{C_R}{\Delta}, \quad \mu_R = \frac{L_R}{\Delta}, \tag{1}$$

$$\varepsilon_L = -\frac{1}{\omega^2 L_L \Delta}, \quad \mu_L = -\frac{1}{\omega^2 C_L \Delta} \tag{2}$$

where $\Delta$ is the length of the unit cell in the $x$ and $y$ direction, and the subscripts "R" and "L" refer to the case of mimicking a DPM and DNM respectively.

In order to mimic the invisible gateway device in Figure 1 by using the unit cells of L-C network in Figure 2a and 2b, two requirements should be met at the design frequency. One is that both the DPM and DNM network must behave like effective media with $\sqrt{\mu_R \varepsilon_R} f \Delta \leq 1$ and $\sqrt{\mu_L \varepsilon_L} f \Delta \leq 1$. The other is $\varepsilon_L = -\varepsilon_R$ and $\mu_L = -\mu_R$.

Here we choose the unit cell parameters as $\Delta$=6mm, $L_R$=47nH, $C_R$=82pF to meet all the above requirements in a design frequency $f_0$=51MHz. These parameters result in an effective wavelength of 60mm. The experimental device is illustrated in Figure 2c. It is fabricated on a grounded flame retardant 4 (FR4) substrate of thickness 1mm and dielectric constant $\varepsilon_r$=4.3. The whole structure has 61 grid nodes in the $x$ direction and 41 grids in $y$ direction. The distance between two adjacent nodes is equal to the length of a unit cell $\Delta$=6mm (i.e., $0.1\lambda$). The structure measures totally about 390mm×270mm. The magnified views of the grid nodes in different regions are also shown in the insets of the Figure 2c. The node in the DPM region consists of four surface-mounted inductors in series and one capacitor in shunt to the ground by a via-hole (see schematic plot in Figure 2a). The left-handed unit cell consists of four surface-mounted capacitors in series and one inductor in shunt to the ground by a via-hole (see the schematic plot in Figure 2b). The grounded PEC plates are used to mimic the PEC walls in Figure 1. In our design, the outer boundaries of the DPM regions are truncated by being connected with Bloch impedances [24] which are applied to achieve matching absorption to mimic the infinitely extended background. A topology which is very similar to the Yee grids in the finite-difference time-domain (FDTD) method is employed at the interface of the DPM region and its complementary counterpart [25]. As shown in the lower inset of Figure 2c, the series branches of the node consists of the parallel connection of $2L_R$ and $C_L/2$, while the shunt branche consists of the parallel connection of $C_R/2$ and $2L_L$. With such a topology, a more precise equivalence to the continuity condition of tangential electric and magnetic field at the interface between two different medium can be achieved.

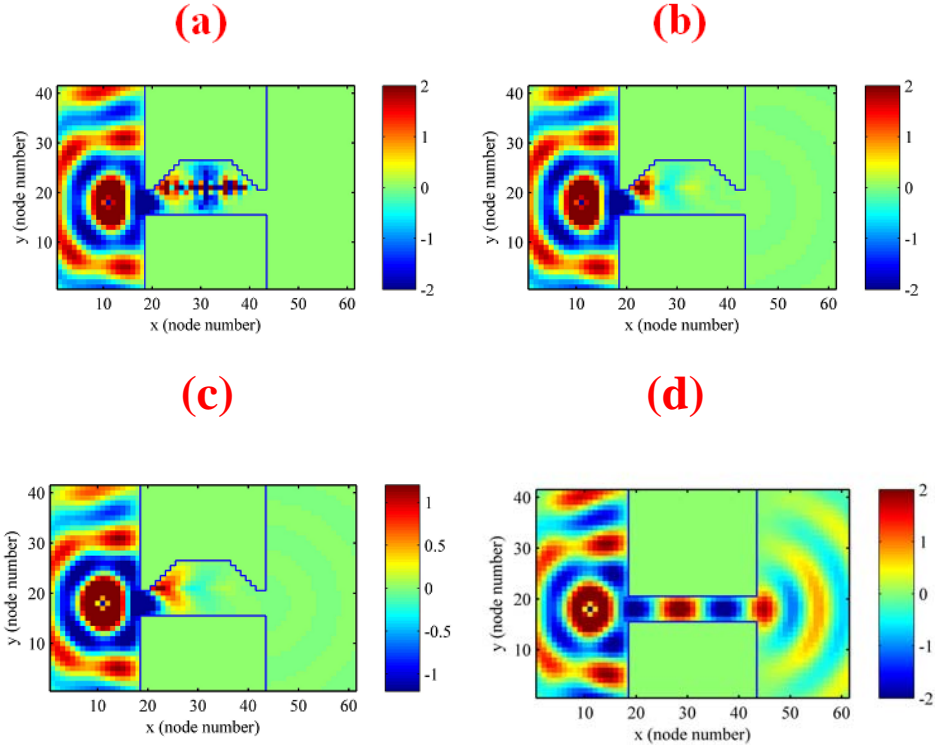

Figure 3 | simulated node voltage distribution. Invisible gateway with DNM (a) for lossless case, (b) of very little loss ($Q = 10^5$), (c) $Q = 10$; (d) with PEC instead of DNM.

Before fabricating the device and performing the experiment, we have evaluated the performance of the device using numerical simulations. We generated the whole network in Figure 2c using Agilent's Advanced Design System (ADS). In the simulation, a 1-A current source is connected between the center of the node (11, 18) and the ground to generate the point source excitation. One advantage of the simulation is that we can study the ideal situation when all the inductors and capacitors are lossless. The simulated result for lossless case at the design frequency is shown in Figure 3a. Strong surface waves can be clearly observed at the interface of the DPM and DNM with symmetrical pattern. The waves from the point source are completely blocked from propagating to the right domain. When a small loss is introduced (the quality factor for all the inductors and capacitors is set to be $Q = 10^5$), the pattern of the surface waves changes conspicuously as compared to the lossless case, as shown in Figure 3b. This means that the resonance in the region near the interface is sensitive to the loss, as expected. However, the field in the right domain is still very weak which means that most of the waves excited by the source in the left domain are mostly blocked by the gateway device. This phenomenon can also be observed when $Q$ is set to be 10 as another example, as shown in Fig.3(c). This means the illusion property of the invisible gateway is robust against material loss. This improves the possibility of getting reasonably good results in experiment where loss is inevitable. For comparison, Figure 3d shows the simulated voltage distribution when the DNM in the trapezoidal region is replaced by PEC. In this case, the waves can evidently propagate to the right domain.

To investigate the realistic performance of the present invisible gateway device, an Agilent E5071C vector network analyzer (VNA) was employed to take transmission coefficient measurement, which is in proportional to the voltage of the grid nodes. Port 1 of the VNA provides the excitation via a coaxial feed, with its outer conductor mounted onto the ground plane at the backside of the FR4 substrate, and its center pin extending through a hole in the substrate and soldered to the center of the node (11, 18), which is same as that in the simulation. Hence, a point source is introduced to excite a cylindrical wave. Port2 of the VNA provides a near-field coaxial probe that can scan over the surface of the whole structure. To ensure the precision of the measurement, the disturbance of the voltage distribution as the detecting probe touching the node must be well controlled. Here, a broadband amplifier with very high input impedance is designed and inserted between the detecting probe and Port2 of the VNA. The amplifier is designed based on the topology of differential amplifier with current feedback. The measured input impedance is more than 2k Ohms and the measured voltage gain is 20dB$\pm$0.2dB over the frequency band from DC to 100MHz.

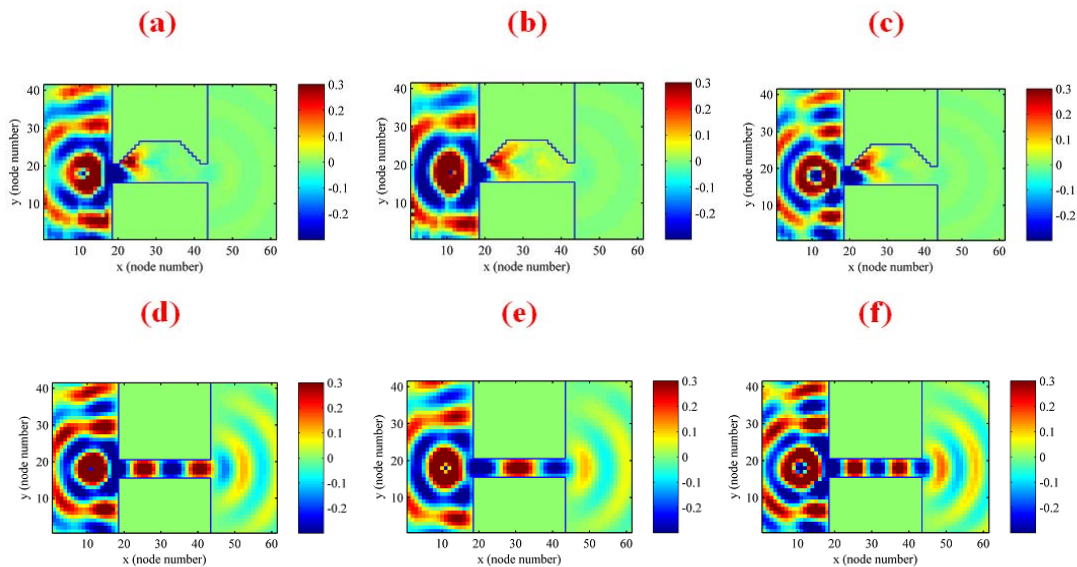

Figure 4 | Measured node voltage distribution. Gateway with DNM at the trapezoidal region at frequencies (a) f=51MHz, (b) 46MHz, and (c) 56MHz; Gateway with DNM replaced by PEC at frequencies (d) f=51MHz, (e) 46MHz, and (f) 56MHz.

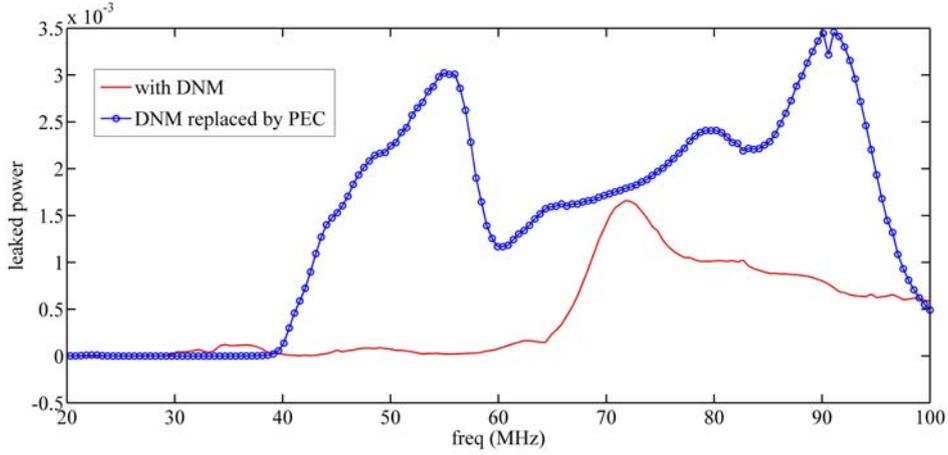

Figure 5 | The leaked power from the left domain for the case of DNM and PEC. The data is derived from the experimental results.

The measured results at the design frequency f=51MHz is shown in Figure 4a. For comparison, Figure 4d shows the measured voltage distribution when the DNM is replaced by the PEC. These two results are very similar to the numerical simulations shown in Figure 3c and 3d, which verifies that our device has the realistic illusion property of the invisible gateway. In the frequency band around 51MHz (the designed frequency), the chip inductors and capacitors have non-resonant nature and the values of $L_R$, $L_L$ and $C_R$, $C_L$ change gradually when the frequency changes. Although the requirement of the complementary medium condition is only satisfied strictly at the designed frequency theoretically, the "optical cancellation" property still remains effective even if the material properties are slightly off. The reasonably wide functional bandwidth is evident in Figures 4 and Figure 5. Figure 4b and 4c show the functionalities of the device with DNM at frequencies 46MHz and 56MHz, while Figure 4e and 4f show the functionalities of the device with PEC instead of DNM at frequencies 46MHz and 56MHz. It is seen that, the illusion property can clearly be extended to such frequencies. To quantitatively evaluate the bandwidth of our device, we calculated the power leaked into the right domain for the case with DNMs, $P_{DNM}$ and with PECs instead of DNMs, $P_{PEC}$ based on

$$P = \sum_i V(i) \cdot I_{out}^*(i), \qquad (3)$$

where $V(i)$ and $I_{out}(i)$ are the node voltages and outward currents on the nodes along the right window of the gateway. The leaked power ($P_{DNM}$ and $P_{PEC}$) are plotted in Figure 5 as a function of frequencies. By comparing these two curves, we can state that the device work well from about 35 MHz to 60 MHz.

To summarize, we have designed and fabricated a device to mimic an electromagnetic invisible gateway. The simulation and measurement results confirm its functionalities. In addition, such a device works for a rather broadband. This demonstration will be helpful for future design of an electromagnetic "illusion devices" at microwave frequencies. Other metamaterial illusion optical devices (such as an external cloak) can also be designed and emulated by using the similar transmission-line medium.

## Acknowledgements

This work was supported by the National Natural Science Foundation of China (60990323, and 60990320), and the Knowledge Innovation Program of Chinese Academy of Sciences, Hong Kong RGC grant 600209. Computation resources at Hong Kong were supported by the Shun Hing Education and Charity Fund.